\documentclass[graybox]{svmult}

\usepackage{graphicx}
\usepackage{epstopdf}
\usepackage{mathptmx}       
\usepackage{helvet}         
\usepackage{courier}        
\usepackage{type1cm}        
\usepackage{makeidx}         
\usepackage{graphicx}        
\usepackage{multicol}        
\usepackage[bottom]{footmisc}

\usepackage{latexsym,epsfig,graphicx,subfigure,amsmath,amssymb,amscd,multirow,psfrag,paralist,dsfont,float}

\usepackage{amssymb}
\usepackage{bm}

\makeindex             

\newcommand{\thetavec}{{\boldsymbol{\theta}}}

\newcommand{\thetavechat}{\widehat{\thetavec}}

\newcommand{\gammavec}{{\boldsymbol{\gamma}}}
\newcommand{\Corr}{{\rm Corr}}

\newcommand{\ybar}{{\overline{y}}}

\newcommand{\yvec}{\boldsymbol{y}}

\newcommand{\ti}{R}
\newcommand{\degreeC}{{^\circ\mathrm{C}}}

\makeatletter
\renewcommand\@biblabel[1]{#1.}
\makeatother

\begin{document}

\title*{Statistical Methods for Thermal Index Estimation Based on Accelerated Destructive Degradation Test Data}
\titlerunning{Statistical Methods for Thermal Index Estimation}
\author{Yimeng Xie, Zhongnan Jin, Yili Hong, and Jennifer H. Van~Mullekom}
\authorrunning{Xie, Jin, Hong, and Van~Mullekom}
\institute{Yimeng Xie \at Department of Statistics, Virginia Tech, Blacksburg, VA 24061, \email{xym@vt.edu}
\and Zhongnan Jin \at Department of Statistics, Virginia Tech, Blacksburg, VA 24061, \email{jinxx354@vt.edu}
\and Yili Hong,  corresponding author \at Department of Statistics, Virginia Tech, Blacksburg, VA 24061, \email{yilihong@vt.edu}
\and Jennifer H. Van~Mullekom \at Department of Statistics, Virginia Tech, Blacksburg, VA 24061, \email{vanmuljh@vt.edu}}

\maketitle
\abstract*{Accelerated destructive degradation test (ADDT) is a technique that is commonly used by industries to access material's long-term properties. In many applications, the accelerating variable is usually the temperature. In such cases, a thermal index (TI) is used to indicate the strength of the material. For example, a TI of 200$\degreeC$ may be interpreted as the material can be expected to maintain a specific property at a temperature of 200$\degreeC$ for 100,000 hours. A material with a higher TI possesses a stronger resistance to thermal damage. In literature, there are three methods available to estimate the TI based on ADDT data, which are the traditional method based on the least-squares approach, the parametric method, and the semiparametric method.  In this chapter, we provide a comprehensive review of the three methods and illustrate how the TI can be estimated based on different models. We also conduct comprehensive simulation studies to show the properties of different methods. We provide thorough discussions on the pros and cons of each method. The comparisons and discussion in this chapter can be useful for practitioners and future industrial standards.}


\abstract{Accelerated destructive degradation test (ADDT) is a technique that is commonly used by industries to access material's long-term properties. In many applications, the accelerating variable is usually the temperature. In such cases, a thermal index (TI) is used to indicate the strength of the material. For example, a TI of 200$\degreeC$ may be interpreted as the material can be expected to maintain a specific property at a temperature of 200$\degreeC$ for 100,000 hours. A material with a higher TI possesses a stronger resistance to thermal damage. In literature, there are three methods available to estimate the TI based on ADDT data, which are the traditional method based on the least-squares approach, the parametric method, and the semiparametric method.  In this chapter, we provide a comprehensive review of the three methods and illustrate how the TI can be estimated based on different models. We also conduct comprehensive simulation studies to show the properties of different methods. We provide thorough discussions on the pros and cons of each method. The comparisons and discussion in this chapter can be useful for practitioners and future industrial standards.}

\section{Introduction}\label{sec:introuduction}
\subsection{Background}
Polymeric materials are common in various industrial applications. In current industrial practice, a thermal index (TI) is often used to rate the long-term performance of polymeric materials. As specified in industrial standard UL~746B~\cite{UL746B}, the TI of a polymeric material can be considered as a measure of the material's ability to retain a specific property (e.g., physical or electrical properties) under exposure to elevated temperatures over a prolonged period of time (e.g., 100{,}000 hours). The interpretation of the TI is as follows. A material with a TI value of $200\degreeC$ is expected to maintain the specific property for exposure to a temperature of $200\degreeC$ for 100,000 hours. Thus, a material with a higher TI rating is expected to demonstrate a stronger resistance to thermal exposure as compared to those with a lower TI ratings. The TI can also be used to determining suitability for a particular application, and for comparing multiple materials. When a material is introduced to a field, its TI can be compared to a list of similar materials with known TI values, which can give insights for the long term performance of the new material. Therefore, estimating the TI for a material is an important task in evaluating material performance.

To estimate the TI, data which track the material property over time need to be collected. Such data are referred to as degradation data. However, the degradation of the material performance is often gradual and can take years to observe deteriorations. To collect information in a timely manner, accelerated degradation test (ADT) is often used. In the setting of TI estimation, temperature is the accelerating variable. When measuring the material performance, such as the tensile strength, the sample will be stretched until it breaks. Because the sample is destroyed in the testing procedure, only one data point can be collected from one sample. Such type of ADT is called accelerated destructive degradation testing (ADDT). Due to the nature of the testing,  ADDT is a commonly used technique for evaluating long-term performance of polymeric materials. Examples of ADDT data includes the Adhesive Bond B data in \cite{Escobaretal2003}, the Polymer data in \cite{Tsaietal2013}, the Seal Strength data in \cite{LiDoganaksoy2014}, and the Formulation K data in \cite{XieKingHongYang2015}.

To use the ADDT data for the TI estimation, a statistical method is needed. In literature, there are three methods available to estimate the TI based on ADDT data, which are the traditional approach based on the least-squares method, the parametric approach based on maximum likelihood (ML) method, and the semiparametric approach based on splines method. The traditional procedure is the one that is currently specified in the industrial standards UL~746B~\cite{UL746B}, which is commonly used  to evaluate material in applications. The traditional approach is a two-step approach using polynomial fittings and least-squares methods. In the statistical literature, the parametric method is also commonly used to model the ADDT data, and the ML method is used for parameter estimation. Recently, a semiparametric method is proposed to analyze ADDT data in Xie et al.~\cite{XieKingHongYang2015}. The basic idea of the semiparametric method is to use monotonic splines to model the baseline degradation path and use a parametric method to model the effect of accelerating variable.

The objective of this chapter is to provide a comprehensive review of the three methods and illustrate how the TI can be estimated based on different models. We also conduct comprehensive simulation studies to show the properties of different methods. Then, we provide thorough discussions on the pros and cons of each method. The comparisons and discussions in this chapter can be useful for practitioners and future industrial standards.

\subsection{Related Literature}
Degradation data were used to access products and material reliability in early work such as Nelson~\cite[Chapter 11]{Nelson1990}, and Lu and Meeker~\cite{LuMeeker1993}. There are two types of degradation data: repeated measures degradation test (RMDT) data and ADDT data. For RMDT data, multiple measurements can be taken from the same unit. For ADDT data, only one measurement can be taken from the same unit, due to the destructive nature of the measuring procedure. Different types of methods are used to analyze RMDT and ADDT data. The majority of the degradation literature is on RMDT data analysis, features two major classes of models: the general path model (e.g., \cite{MeekerEscobar1998} and \cite{Hongetal2015}) and stochastic process models (e.g., \cite{Whitmore1995}, \cite{ParkPadgett2005}, and \cite{WangXu2010}). A review of statistical degradation models and methods are available in Meeker, Hong, and Escobar~\cite{MeekerHongEscobar2011}, and Ye and Xie~\cite{YeXie2015}.

This chapter focuses on the analysis of ADDT data and their corresponding TI estimation procedures. Regarding ADDT analysis, the traditional approach for TI estimation using the least-squares method is described in~UL~746B~\cite{UL746B}. Parametric models are quite common in ADDT analysis, for example, in Escobar et al.~\cite{Escobaretal2003},  Tsai et al.~\cite{Tsaietal2013}, and Li and Doganaksoy~\cite{LiDoganaksoy2014}. King et al.~\cite{Kingetal2016} applied both the traditional and parametric approaches to ADDT data analysis and TI estimations. King et al.~\cite{Kingetal2016} also did a comprehensive comparisons for the two approaches in TI estimations. Xie et al.~\cite{XieKingHongYang2015} developed a semiparametric approach for ADDT data analysis, in which the monotonic splines are used to model the baseline degradation path and the Arrhenius relationship is used to describe the temperature effect. However, the TI estimation procedure was not developed in \cite{XieKingHongYang2015}.

In this chapter, we develop the TI estimation based on the semiparametric method after providing a review of the existing methods in TI estimations. We also conduct a comprehensive simulations to compare the three methods. In terms of software implementation, Hong et al. \cite{Raddt} implements the three methods and their corresponding TI estimation procedures into an R package ``ADDT''. Details and illustrations of the R package ADDT is available in Jin et al.~\cite{JinXieHongVanMullekom2017}.

\subsection{Overview}
The rest of this chapter is organized as follows. Section~\ref{sec:AT.TI} introduces the concept of ADDT, examples of ADDT data, and the concept of TI. Section~\ref{sec:three.methods} presents the three different methods that can be used to model ADDT data and their corresponding procedures for TI estimation. The three different methods are the traditional methods, the parametric method, and the semiparametric method. Section~\ref{sec:simulation} conducts extensive simulations to compare the performances of the estimation procedures. Section~\ref{sec:discussions} provides a comprehensive discussion on the pros and cons of each method and suggestions for practitioners.

\section{Accelerated Tests and Thermal Index}\label{sec:AT.TI}
In this section, we give a more detailed introduction to ADDT and TI.

\subsection{Test Plans}\label{sec:ADDT.intro}
The test plan of an ADDT consists of the temperature levels, the measuring time points, and the number of samples allocated to each combination of the temperature levels and measuring time points. Table~\ref{tab:test.plan} illustrates a test plan for an ADDT. Four elevated temperature levels are considered in the test, which are $250\degreeC$, $260\degreeC$, $270\degreeC$, and $280\degreeC$. There are five measuring time points considered in this plan, which are 552 hours, 1008 hours, 2016, hours, 3528 hours, and 5040 hours. At the initial time (time zero), there are ten sample units tested under the normal temperature level to serve as the baseline. For each combination of temperature level and time points, there are five sample units tested to obtain the measurements for the material property. To measure some properties like tensile strength, the unit will be destroyed after the measurement. Note that equal sample allocation is used in Table~\ref{tab:test.plan}. However, unequal sample size allocation is also seen in practice. See King et al.~\cite{Kingetal2016} for more detailed discussion on the test plans.

\begin{table}
\begin{center}
\caption{Illustration sample size allocation for an ADDT.}\label{tab:test.plan}
\begin{tabular}{ccccccc}\hline\hline
Temperature  & \multicolumn{6}{c}{Measuring Points (Hours)} \\ \cline{2-7}
($\degreeC$) &0 & 552 & 1008 & 2016 & 3528 & 5040 \\ \hline
-   &   10 & &   &   &  &    \\
250 &      & 5 & 5 & 5 & 5& 5  \\
260 &      & 5 & 5 & 5 & 5& 5  \\
270 &      & 5 & 5 & 5 & 5& 5  \\
280 &      & 5 & 5 & 5 & 5& 5  \\ \hline\hline
\end{tabular}
\end{center}
\end{table}

\subsection{Data and Notation}\label{sec:data.notation}
\begin{figure}
\begin{center}
\includegraphics[width=.8\textwidth]{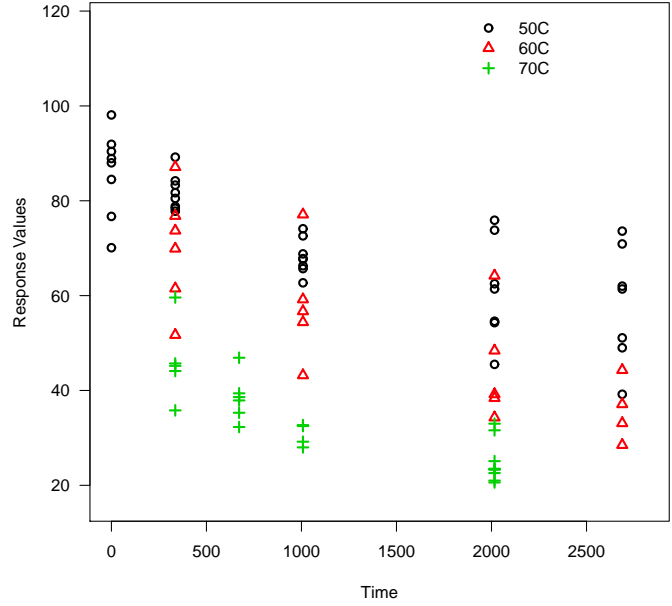}
\end{center}
\caption{Scatter plot of the Adhesive Bond B data. The x-axis is time in hours and the y-axis is strength in Newtons.} \label{fig:AdhesiveBondB}
\end{figure}

The ADDT data record the material property (e.g., the tensile strength of the material) for each unit. Here, we use the Adhesive Bond B example in Escobar et al.~\cite{Escobaretal2003} to illustrate the ADDT data. Figure~\ref{fig:AdhesiveBondB} shows a scatter plot of the Adhesive Bond B data. In general, we observe that there is a decreasing trend over time, while for higher temperature level, the rate of decreasing is faster than those under lower temperature levels.

Here we introduce some notations to the ADDT data that will be necessary for the development of the statistical methods. Let $n$ be the number of temperature levels and $n_i$ be the number of measuring time points for temperature level $i$. The value of the $i$th temperature level is denoted by $A_i$. The corresponding time points are denoted by $t_{ij}$, $j=1, \cdots, n_i$. Note that it is possible that the measure time points are different from different temperature levels. Let $n_{ij}$ be the number samples tested at time $t_{ij}$ for temperature level $i$. Note that the number of samples tested at each time point $t_{ij}$ can also vary. We denote the degradation measurement by $y_{ijk}$ for the $k$th sample at level $i$ of the temperature level $i$ and measuring time $t_{ij}$, $i=1,\cdots, n$, $j=1,\cdots,n_i$, and $k = 1, \cdots, n_{ij}$. The total number of measured samples are $N=\sum_{i=1}^{n}\sum_{j=1}^{n_i}n_{ij}$.

\subsection{Thermal Index}\label{sec:TI.intro}
In this section, we introduce the general concept of the thermal index (TI). In the following, we will use the tensile strength as the interested material property. In a common framework of degradation modeling, the failure time is defined as the first time when the degradation level passes the failure threshold. For example, a failure is said to have occurred when the tensile strength of a sample reaches a certain percentage (e.g., 50\%) of the original tensile strength.

\begin{figure}
\begin{center}
\includegraphics[width=.8\textwidth]{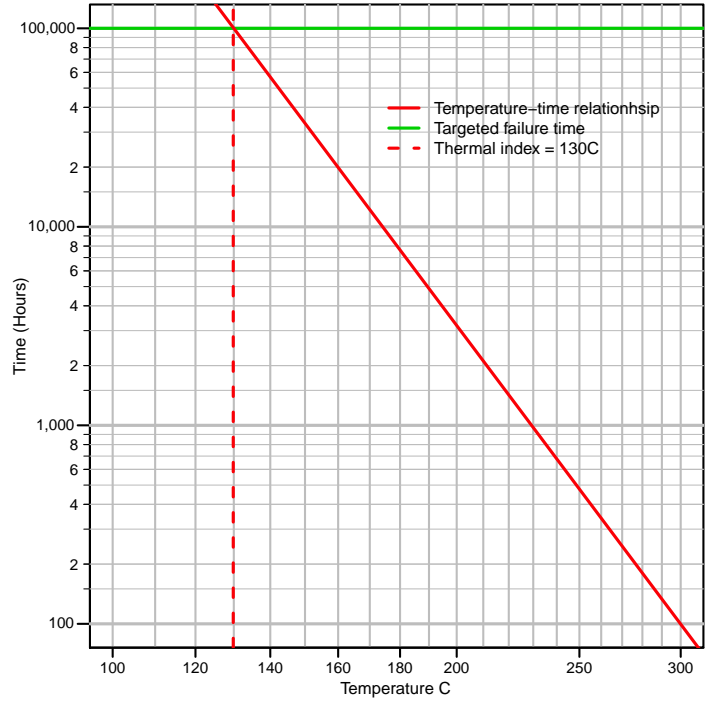}
\end{center}
\caption{Illustration of temperature-time relationship and TI. The x-axis is temperature $A$ on the scale of $1/(A + 273.16)$, and the y-axis is time in hours on base 10 logarithm scale.}\label{fig:illustration.TI.def}
\end{figure}

For degradation processes that are accelerated by temperature, the Arrhenius relationship is widely used to model the relationship between the degradation and temperature. In particular, the Arrhenius model uses the following transformed temperature,
\begin{align}\label{eqn:Arrhenius.model}
h(A)=\frac{-11605}{A + 273.16},
\end{align}
where $A$ is the temperature value in degrees Celsius, the constant 11605 is the reciprocal of the Boltzmann's constant (in units of eV). Note that the constant 273.16 is for converting the Celsius temperature scale to the Kelvin temperature scale. For the convenience of modeling, we define,
\begin{align*}
x=\frac{1}{A + 273.16}, \quad\textrm{and} \quad x_i=\frac{1}{A_i + 273.16}.
\end{align*}
Through the modeling of the degradation data, which will be detailed in Section~\ref{sec:three.methods}, the mean time to failure at $x$ can be described by a relationship $m(x)$. For targeted time to failure $t_d$ (e.g., $t_d=100{,}000$ hours), the corresponding temperature level $\ti$ can be obtained by solving $x_d$ from $m(x_d)=t_d$. Because $$x_d=m^{-1}(t_d)=\frac{1}{\ti + 273.16},$$ we obtain the corresponding temperature value $\ti$ as
\begin{align}\label{eqn:thermal.index.def}
\ti=\frac{1}{m^{-1}(t_d)}-273.16.
\end{align}
The temperature level $\ti$ in \eqref{eqn:thermal.index.def} is defined as the TI for the material. Figure~\ref{fig:illustration.TI.def} illustrates the temperature-time relationship based on the Arrhenius relationship and the corresponding TI.

Note that the targeted time to failure is not required to be fixed at $100{,}000$ hours. For example, if there is an existing material with a known TI (e.g., 220$\degreeC$), its targeted time to failure $t_d^{\textrm{old}}$ can be obtained. For a new material, its TI can be obtained by using $t_d^{\textrm{old}}$ as the targeted time. In this case, the TI for the new material is called the relative TI because it compares to an existing material, see King et al.~\cite{Kingetal2016} for more details.

\section{Statistical Methods for Thermal Index Estimations}\label{sec:three.methods}
This section covers the statistical methods for the TI estimation. We first review the traditional and the parametric methods as described in King et al.~\cite{Kingetal2016}. Then, we derive the TI estimation based on the semiparametric model in Xie et al.~\cite{XieKingHongYang2015}.

\subsection{The Traditional Method}\label{sec:traditional.method}
The traditional method is the methodology that is described in UL~746B~\cite{UL746B}, which is the currently accepted standard for ADDT data analysis in industry. The procedure essentially is a two-step approach. The basic idea is to find an appropriate model to link the time to failure to the level of degradation, and then estimate the parameters by applying the least-squares technique. The estimated failure time is done by interpolating the fitted curves. If there is no material-specific knowledge on the degradation relationship, the UL standards recommend using a third-order polynomial fitting.

Specifically, for temperature level $i$, one first computes the points $\{t_{ij}, \ybar_{ij\cdot}\}$, $j=1, \cdots, n_i$, where
$$
\ybar_{ij\cdot}=\frac{1}{n_{ij}}\sum_{k=1}^{n_{ij}}y_{ijk}
$$
is the average of the batch of observations at time $t_{ij}$ for temperature level $i$. A third order polynomial $a_{0i}+a_{1i}t+a_{2i}t^2+a_{3i}t^3$ is used to fit the data points $\{t_{ij}, \ybar_{ij\cdot}\}$, $j=1, \cdots, n_i$, separately for each temperature level. Here, $(a_{0i}, a_{1i}, a_{2i}, a_{3i})'$ are the polynomial coefficients to be estimated by the least-squares method.

After obtaining the estimates of $(a_{0i}, a_{1i}, a_{2i}, a_{3i})'$, the mean failure time $m_i$ for temperature level $i$ can be obtained through interpolation. In particular, one needs to solve,
\begin{align}\label{eqn:tf.poly.intp}
a_{0i}+a_{1i}m_i+a_{2i}m_i^2+a_{3i}m_i^3=y_{f},
\end{align}
where $y_f$ is the failure threshold. The failure threshold is usually set at 50\% of the initial strength, though different values may be set according to the specification of different applications.

Through the polynomial interpolation, a set of data points $\{x_i, m_i\}, i=1,\cdots, n$ are obtained where $x_i$ is the transformed temperature as defined in \eqref{eqn:Arrhenius.model}. The least-squares method is used again to fit a straight line to data points $\{x_i, \log_{10}(m_i)\}, i=1,\cdots, n$. That is, to fit the following model,
\begin{align*}
\log_{10}(m_i)=\beta_0+\beta_1x_i, i=1,\cdots, n,
\end{align*}
to obtain the estimates of $\beta_0$ and $\beta_1$. Note that the base 10 logarithm is used here because it is more popular in engineering literature. In the traditional method the temperature-time relationship is represented as
\begin{align}\label{eqn:lsa}
\log_{10}[m(x)]=\beta_0+\beta_1x.
\end{align}
With the fitted temperature-time relationship in~\eqref{eqn:lsa}, the TI based on the traditional method is obtained as
\begin{align}\label{eqn:lsa.ti}
R=\frac{\beta_1}{\log_{10}(t_d)-\beta_0}-273.16\,.
\end{align}
where $t_d$ is the target time and $t_d=100{,}000$ is often used.

The traditional method is fairly intuitive and straightforward to compute, which is of advantages. Here we provide some other considerations for the traditional method. The interpolation based method requires the degradation level to reach the failure threshold so that $m_i$ can be obtained for level $i$. Otherwise, all data collected at level $i$ can not be used for analysis. The number of temperature levels for the ADDT is usually small (i.e., around 4). Thus only a few number of observations are available to fit the model in \eqref{eqn:lsa}. Furthermore, the two-step approach detailed in the current standard does not specify a method to quantify the uncertainty associated with the TI estimation. For ADDT data, one would expect higher temperature levels to yield shorter lifetimes. Due to randomness in the data and the flexibility of polynomials, the traditional method can produce estimated failure times that are not monotonically increasing with temperature, which might be unrealistic. With parametric models, most of the concerns can be addressed.

\subsection{The Parametric Method}\label{sec:para.method}
In statistical literature, parametric methods are prevalent in ADDT data analysis such as in \cite{Escobaretal2003}, \cite{Tsaietal2013}, and~\cite{LiDoganaksoy2014}. In the parametric method, the primary method for estimation and inference is based on a parametric model and maximum likelihood theory. Here, we give a brief description for the parametric method summarized in King et al.~\cite{Kingetal2016}.

In this setting, the parametric model for the degradation measurement is represented as
\begin{align}\label{eqn:degradation.path.model}
y_{ijk}&=\mu(t_{ij}; x_i)+\epsilon_{ijk},
\end{align}
where $\mu(t_{ij}; x_i)$ is the underling degradation path and $\epsilon_{ijk}$ is an error term. Since the tensile strength is decreasing over time, the function $\mu(t; x_i)$ is specified as a decreasing function of $t$. Consequently, a higher temperature usually lead to a higher rate of degradation. The function $\mu(t; x_i)$ is also a decreasing function of the temperature.

For a specific $x$, the mean time to failure $m(x)$ can be solved from
$$\mu[m(x); x]=y_f,$$
leading to the temperature-time relationship as
$$m(x)=\mu^{-1}(y_f; x).$$
The TI can be solved from $m(x_d)=t_d$, which is equivalent to solve $x_d$ from
$$\mu(t_d; x_d)=y_f.$$
The TI can be computed from the solution $x_d$. That is,
$$
\ti=\frac{1}{x_d}-273.16.
$$

To proceed with the modeling, one needs to be specific about the form of $\mu(t,x)$. For polymer materials, the parametric form in \cite{VacaTrigoMeeker2009} is often used. In particular,
\begin{align}\label{eqn:mu.fun}
\mu(t;x)=\frac{\alpha}{1+\left[\dfrac{t}{\eta(x)}\right]^{\gamma}},
\end{align}
where $\alpha$ is the initial degradation level, $\eta(x)=\exp(\nu_0+\nu_1x)$ is the scale factor based on the Arrhenius model, and $\gamma$ is the shape parameter determining the steepness of the degradation path.

Let $p=y_f/\alpha$ be the proportion of decreasing for the failure threshold from the initial degradation level. Based on the model in \eqref{eqn:mu.fun}, the mean time to failure at $x$, $m(x)$, is obtained by solving $\mu[m(x); x]=p\alpha$. Specifically, the temperature-time relationship is
\begin{align}\label{eqn:log.tf.mla.p.nu}
\log_{10}[m(x)]=\beta_0+\beta_1x,
\end{align}
where
$$\beta_0=\frac{\nu_0}{\log(10)}+\frac{1}{\gamma\log(10)}\log\left[\frac{1-p}{p}\right], \quad \textrm{and}\quad \beta_1=\frac{\nu_1}{\log(10)}.$$
When $p=1/2$, $\beta_0$ reduces to $\nu_0/\log(10)$. The TI at $t_d$ can be computed as
\begin{align}\label{eqn:lsa.ti.mle}
\ti=\frac{\beta_1}{\log_{10}(t_d)-\beta_0}-273.16.
\end{align}

The model in \eqref{eqn:degradation.path.model} is estimated by ML method. The error term is modeled as
\begin{align}\label{eqn:error.dist}
\varepsilon_{ijk} \sim \textrm{N}(0, \sigma^2),\quad\textrm{and}\quad \Corr(\varepsilon_{ijk}, \varepsilon_{ijk'})=\rho,\,\, k \ne k'.
\end{align}
The parameter $\rho$ represents the within-batch correlation. The unknown parameters is denoted by $\thetavec=(\nu_0, \nu_1, \alpha, \gamma, \sigma, \rho)'$. The likelihood is
\begin{align}\label{eqn:likelihood}
L(\thetavec)&= \prod_{i,j}(2\pi)^{-\frac{n_{ij}}{2}} |\Sigma_{ij}|^{-\frac{1}{2}}\exp{\left\{-\frac{1}{2}[\yvec_{ij}-\boldsymbol{\mu}(t_{ij},x_i)]'\Sigma_{ij}^{-1}[\yvec_{ij}-\boldsymbol{\mu}(t_{ij},x_i)]\right\}},
\end{align}
where $\yvec_{ij}=(y_{ij1},\cdots,y_{ijn_{ij}})'$ is the corresponding vector of degradation measurements which follows the multivariate normal distribution with mean vector $\boldsymbol{\mu}(t_{ij};x_i)$, an $n_{ij}\times 1$ vector of $\mathbf{\mu}(t_{ij}; x_i)$'s, and covariance matrix $\Sigma_{ij}$, an $n_{ij}\times n_{ij}$ matrix with $\sigma^2$ on the diagonal entries and $\rho\sigma^2$ on the off-diagonal entries. The parameter estimates $\thetavechat$ are obtained by maximizing \eqref{eqn:likelihood}. The estimate of $\ti$ is obtained by evaluating~\eqref{eqn:lsa.ti.mle} at the estimate $\thetavechat$.

The parametric model can overcome the shortcoming of the traditional method, and allows for statistical inference. However, for the parametric method, one needs to find an appropriate form for $\mu(t_{ij}; x_i)$.

\subsection{The Semiparametric Method}\label{sec:semi.para}
Xie et al. \cite{XieKingHongYang2015} proposed the following semi-parametric functional forms for $\mu(t_{ij}; x_i)$ for the model in \eqref{eqn:degradation.path.model}. That is,
\begin{align}\label{eqn:semi-parametric degradation model}
&\mu(t_{ij}; x_i) = g\left[\eta_{i}(t_{ij};\beta);\gammavec\right],  \\\label{eqn:scale.acc}
&\eta_{i}(t; \beta) =\frac{t}{\exp(\beta s_i)},\quad s_i =x_i-x_{\max}.
\end{align}
Here, $g(\cdot)$ is a monotonic decreasing function with parameter vector $\gammavec$, and $\beta$ is the parameter for the temperature effect. The quantity
$$x_{\max}=\frac{1}{\max_{i}\{A_i\} + 273.16}$$
is the transformed value of the highest level of temperature. At the highest temperature level, $s_{\max} =x_{\max}-x_{\max}=0,$ then
$$
\mu(t; x_{\max}) = g(t;\gammavec).
$$
Thus, the function $g(\cdot)$ is interpreted as the baseline degradation path. The advantage of using the maximum temperature level as the baseline is that its degradation level will reach the failure threshold in most ADDT. The $g(\cdot)$ is constructed nonparametrically by monotonic splines, which is the nonparametric component of the model. The use of the monotonic splines retains the physical meaning of the degradation mechanism (i.e., monotonicity), and it is also flexible because one does not need to find a parametric form for the degradation paths. The Arrhenius model is used for describing the acceleration effect, which is the parametric component of the model. Thus, the model in \eqref{eqn:semi-parametric degradation model} is called a semiparametric model.

The distribution of the error terms $\varepsilon_{ijk}$ are specified in \eqref{eqn:error.dist}. Let $\thetavec=(\gammavec', \beta, \sigma, \rho)'$ be the vector containing all unknown parameters. The estimation of $\thetavec$ is through an iterative procedure that maximizes the loglikelihood function. The details of monotonic spline construction and parameter estimation are referred to Xie et al.~\cite{XieKingHongYang2015}.

Here we derive the TI estimation based on the semiparametric model in \eqref{eqn:semi-parametric degradation model}. Let $g_0=g(0)$ be the initial degradation level and $p$ be the proportion reducing from the initial degradation (i.e., $p=y_f/g_0$). The mean time to failure for the temperature level $x$ is denoted by $m(x)$, which can be solved from
$$
g\left[\frac{m(x)}{\exp[\beta(x-x_{\max})]}\right]=pg_0.
$$
We obtain the temperature-time relationship as,
$$
m(x)=g^{-1}(pg_0)\exp[\beta(x-x_{\max})],
$$
which is equivalent to
\begin{align}
\log_{10}[m(x)]=\beta_0+\beta_1x.
\end{align}
Here, $$\beta_0=\log_{10}[g^{-1}(pg_0)]-\frac{\beta x_{\max}}{\log(10)},\quad \textrm{and}\quad \beta_1=\frac{\beta}{\log(10)}.$$
The TI is computed as,
\begin{align}\label{eqn:semipara.ti}
\ti=\frac{\beta_1}{\log_{10}(t_d)-\beta_0}-273.16.
\end{align}
The estimates of the TI $\ti$ can be obtained by substituting the estimate of $\thetavec$ into~\eqref{eqn:semipara.ti}.

\section{An Illustration of Thermal Index Estimation}
In this section, we provide an illustration for the TI estimation using the Adhesive Bond B data in Escobar et al.~\cite{Escobaretal2003}. The computing was done by using the R package ADDT by Hong et al.~\cite{Hongetal2015}.

\subsection{Degradation Path Modeling}
\begin{figure}
\begin{center}
\includegraphics[width=.8\textwidth]{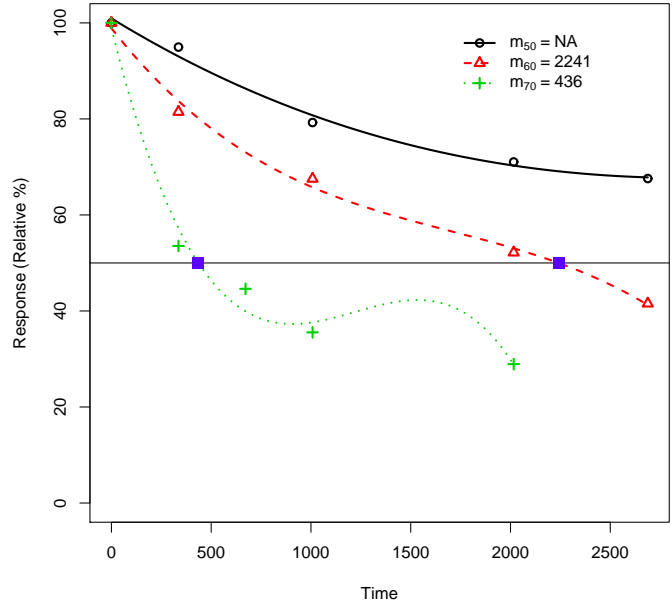}
\end{center}
\caption{Polynomial interpolation for the traditional method for the Adhesive Bond B data. The failure threshold is $p=50\%$.}\label{fig:adhesive.bond.b.poly.intp}
\end{figure}

We apply the traditional method, the parametric method, and the semiparametric method to the Adhesive Bond B data. For the traditional method, Figure~\ref{fig:adhesive.bond.b.poly.intp} shows the polynomial interpolation for the Adhesive Bond B data, when the failure threshold is set to $p=50\%$.
For the temperature level $50\degreeC$, the degradation level has not reached the failure threshold yet. The estimated time to failure $m_{50}$ is not available. Thus, data from this level is discarded from the analysis. In contrast, all data can be used in the parametric and semiparametric methods.

\begin{figure}
\begin{center}
\includegraphics[width=.8\textwidth]{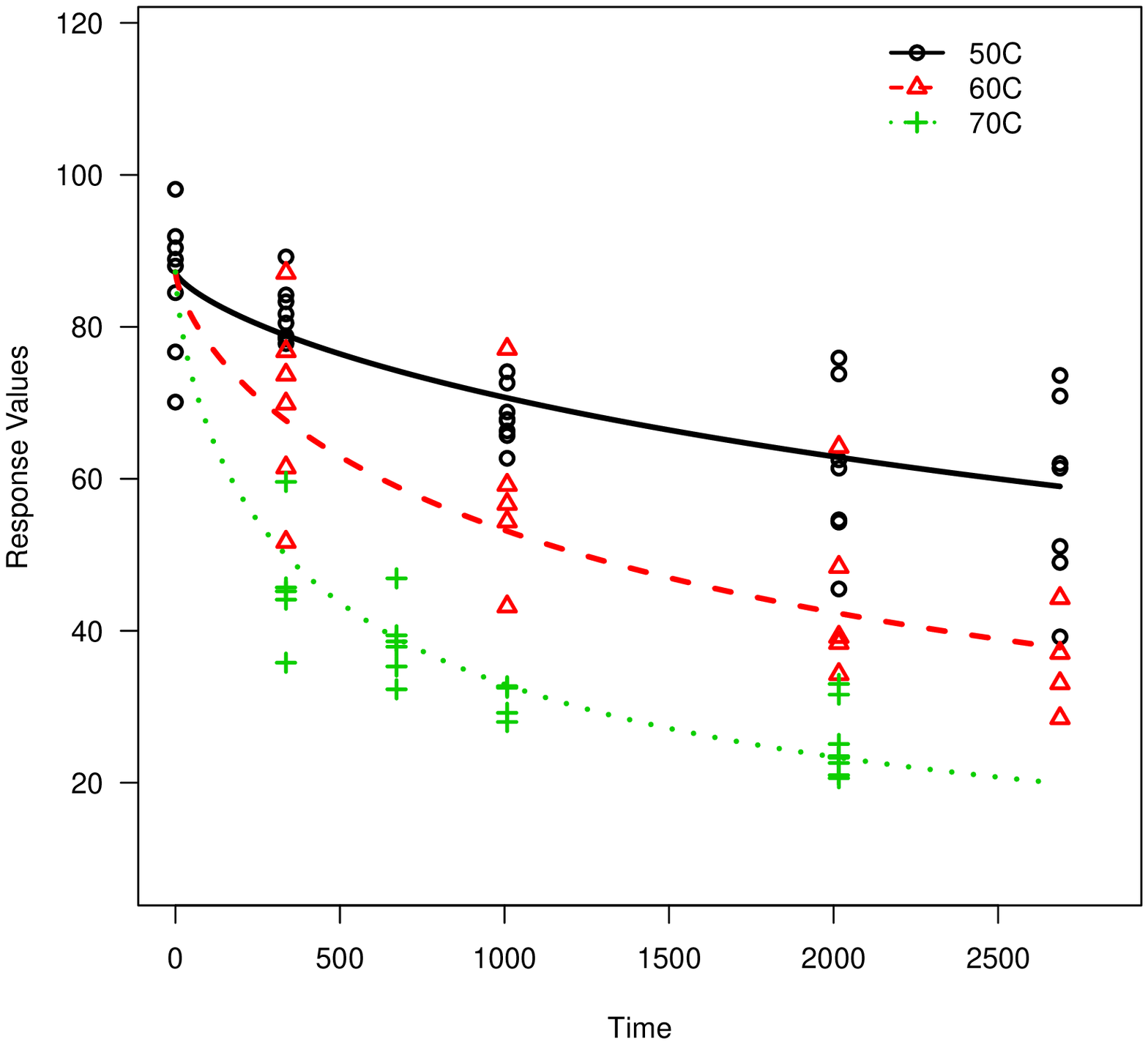}
\end{center}
\caption{Fitted degradation paths using the parametric method for the Adhesive Bond B data. The x-axis is time in hours and the y-axis is strength in Newtons.}\label{fig:adhesive.bond.b.par.fitting}
\end{figure}

\begin{figure}
\begin{center}
\includegraphics[width=.8\textwidth]{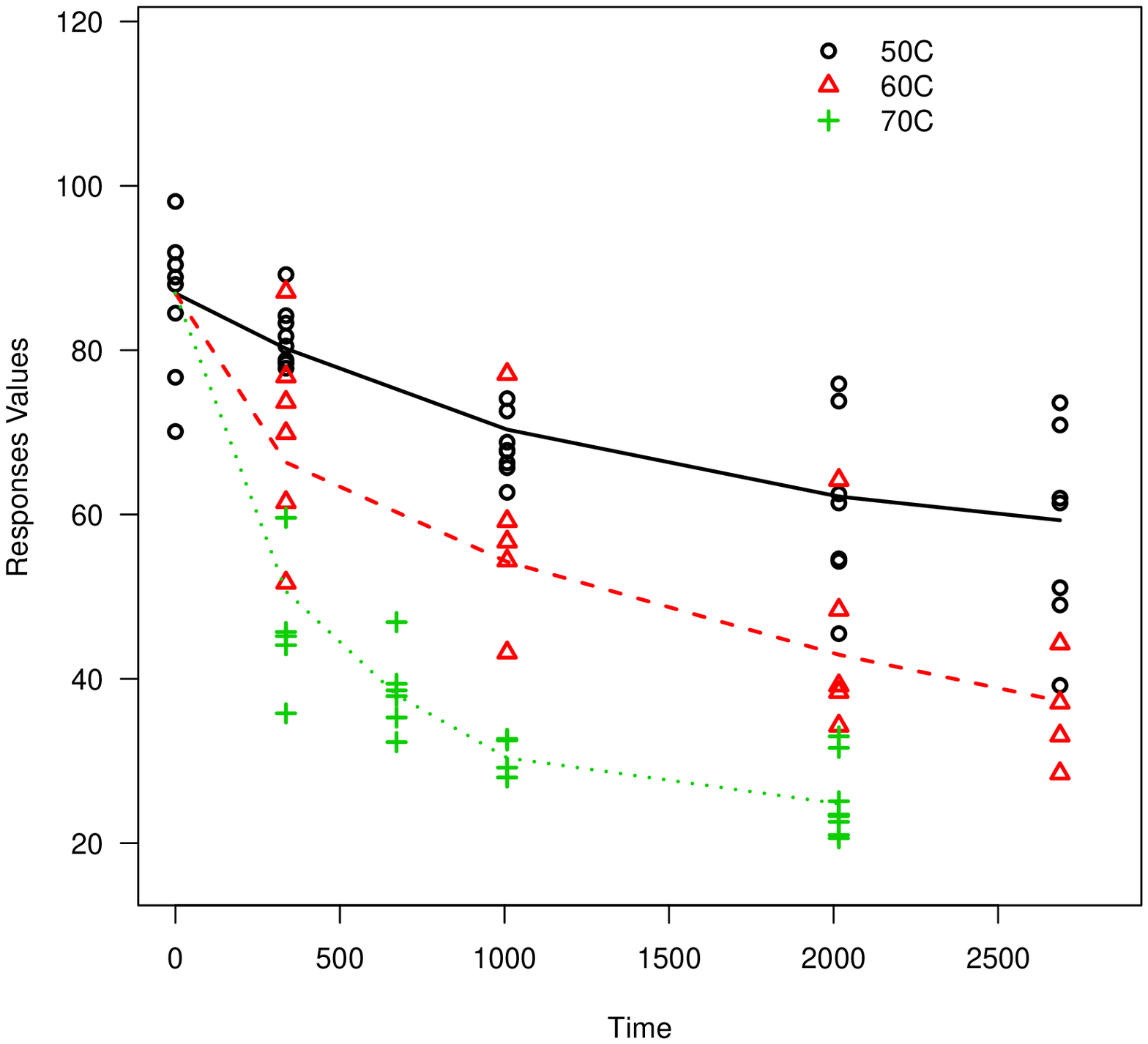}
\end{center}
\caption{Fitted degradation paths using the semiparametric method for the Adhesive Bond B data. The x-axis is time in hours and the y-axis is strength in Newtons.}\label{fig:adhesive.bond.b.semi.par.fitting}
\end{figure}

Figure~\ref{fig:adhesive.bond.b.par.fitting} shows the fitted degradation paths using the parametric method for the Adhesive Bond B data, while Figure~\ref{fig:adhesive.bond.b.semi.par.fitting} shows similar results based on the semiparametric method. Both methods provides good fits to the data. The results in Xie et al.~\cite{XieKingHongYang2015} show that the semiparametric method tends to have a better fit to the degradation data.

\subsection{TI Estimation}
\begin{figure}
\begin{center}
\includegraphics[width=.8\textwidth]{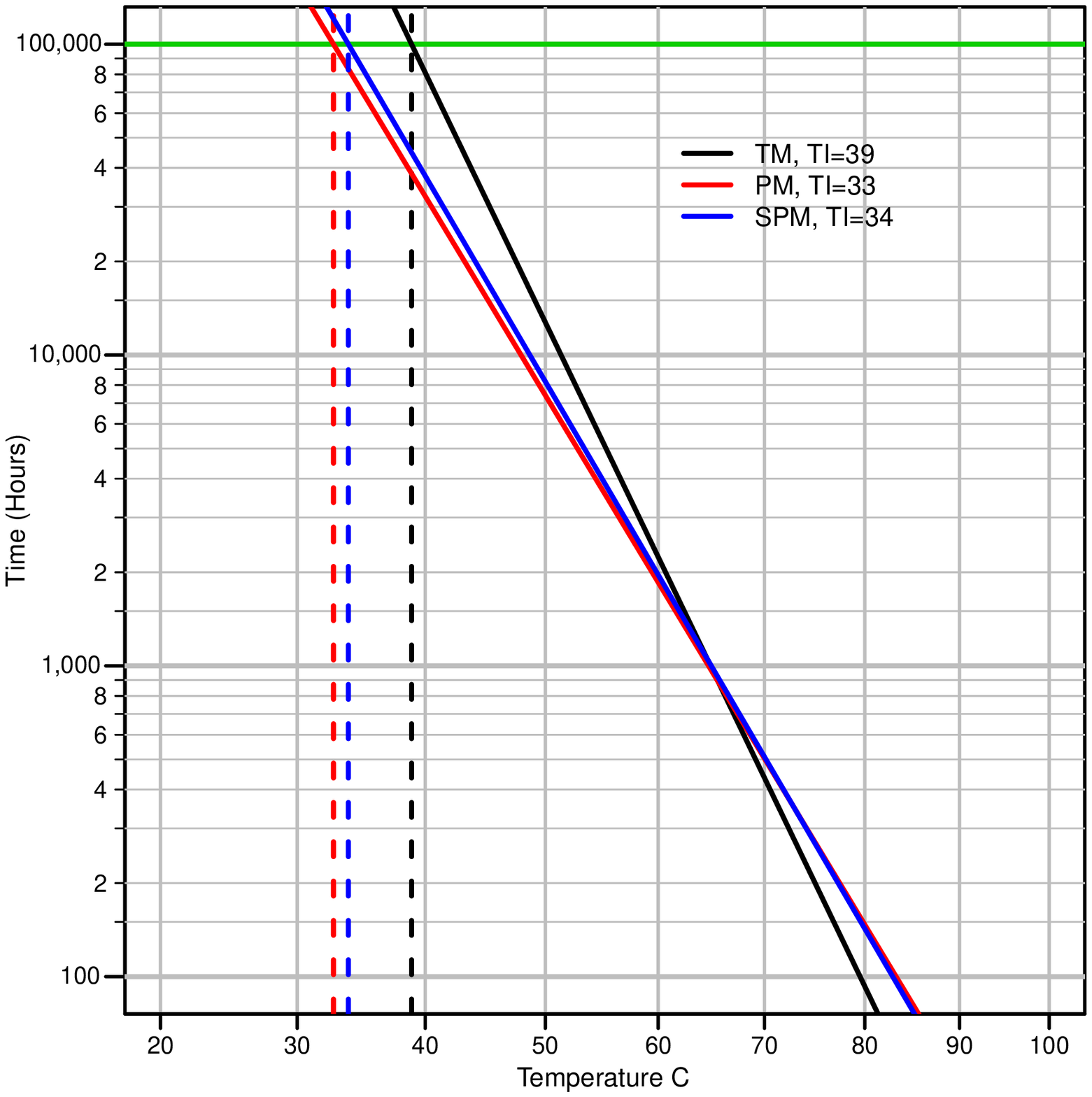}
\end{center}
\caption{Fitted temperature-time relationship lines using the traditional method (TM), the parametric method (PM) and the semiparametric method (SPM), and the corresponding estimated TI for the Adhesive Bond B data.}\label{fig:adhesive.bond.b.TI}
\end{figure}

For illustrations, we compute the TI based on the three methods presented previously. Table~\ref{tab:TI.est} shows the estimated parameters for the temperature-time relationship, and the corresponding TI for the Adhesive Bond B data. In the computing, we use $t_d=100{,}000$ and $p=50\%$. Figure~\ref{fig:adhesive.bond.b.TI} shows the fitted temperature-time relationship lines using the three methods and the corresponding estimated TI for the Adhesive Bond B data. The results based on the parametric method and semiparametric method are quite close to each other, while the results from traditional method is different from these two methods. Section~\ref{sec:simulation} will conduct a simulation study to evaluate the estimation performance.

\begin{table}
\begin{center}
\caption{Estimated parameters for the temperature-time relationship and TI based on the traditional method (TM), the parametric method (PM), and the semiparametric method (SPM) for the Adhesive Bond B data, when $t_d=100{,}000$ and $p=50\%$.}\label{tab:TI.est}
\begin{tabular}{ccccccc}\hline\hline
Methods & & $\beta_0$ & & $\beta_1$ & & TI \\\hline
TM  & & -21.05 & & 8128.4 & & 39 \\
PM  & & -16.18 & & 6480.4 & & 33 \\
SPM & & -16.81 & & 6697.1 & & 34 \\\hline\hline
\end{tabular}
\end{center}
\end{table}

\section{Simulation Studies}\label{sec:simulation}
In this section, simulations are carried out to compare the performance of the traditional method, the parametric method, and the semiparametric method in terms of estimating the TI. We will consider two settings, under which the parametric model is correctly specified, the parametric model is incorrectly specified.

\subsection{Simulation Settings}
For the first setting (Setting I), we generate degradation data from the parametric model in~\eqref{eqn:mu.fun}, and for the parametric method, we still use the same model in~\eqref{eqn:mu.fun} to fit the data, which is corresponding to the case that the model is correctly specified. For model~\eqref{eqn:mu.fun}, the parameter values used in the simulation are $\alpha=9000, \nu_0=-16, \nu_1=12500$, $\gamma=2$, $\sigma=1000$, and $\rho=0$. The failure threshold was set to be $p=50\%$ of the initial degradation level and we used $t_d=100,000$ hours. Under this configuration, the true TI is $\ti=181\degreeC$. The correlation in Setting I is $\rho=0$, to speed up the simulations.

For setting II, we examine model misspecification by generating degradation data from a parametric model that is different from~\eqref{eqn:mu.fun}, but we fit the model in~\eqref{eqn:mu.fun} for the parametric method. In particular, the following model was used to generate data for Setting~II,
\begin{align}\label{eqn:lin.dag}
\mu(t;x)=\alpha\exp\left\{-\left[\dfrac{t}{\eta(x)}\right]\right\},
\end{align}
which was used in Li and Doganaksoy~\cite{LiDoganaksoy2014} to describe the degradation of polymer strength. Here, $\eta(x)=\exp(\nu_0+\nu_1x)$. For model~\eqref{eqn:lin.dag}, the parameter values were set to $\alpha=9000$, $\nu_0=-15.6$, $\nu_1=12471$, $\sigma=1000$, and $\rho=0$. Those values were chosen to match the mean time to failure under 270$\degreeC$ and the true TI $\ti=181\degreeC$ to Setting~I so that the results from both settings are comparable.

For each setting, eight scenarios were considered. For each scenario, we vary the number of time points and the temperature levels. Table~\ref{tab:sim} lists the configuration for each scenario. We considered both four time points and five time points to check the sensitivity to time constraints. The number of temperature levels is from three to five to check the sensitivity to temperature factors. We also considered the range of temperature levels with either higher or lower temperature levels to check the effect of temperature level in terms of distance from use levels. A similar simulation study design was used in King et al.~\cite{Kingetal2016}.

\begin{table}
\caption{The temperature levels and measuring time points for the eight simulation scenarios.}\label{tab:sim}
\begin{center}
\begin{tabular}{c|ccccc|ccccc}\hline\hline
Scenarios & \multicolumn{5}{c|}{Temperature Levels ($\degreeC$)} & \multicolumn{5}{c}{Time Points (Hours)}\\\hline
1: Temp. 3, Time 4& & 250 & 260 & 270 & & 552 & 1008 & 2016 & 3528 & \\
2: Temp. 4, Time 4& & 250 & 260 & 270 & 280 & 552 & 1008 & 2016 & 3528 & \\
3: Temp. 4, Time 4& 240 & 250 & 260 & 270 & & 552 & 1008 & 2016 & 3528 & \\
4: Temp. 5, Time 4& 240 & 250 & 260 & 270 & 280 & 552 & 1008 & 2016 & 3528 & \\\hline
5: Temp. 3, Time 5& & 250 & 260 & 270 & & 552 & 1008 & 2016 & 3528 & 5040 \\
6: Temp. 4, Time 5& & 250 & 260 & 270 & 280 & 552 & 1008 & 2016 & 3528 & 5040 \\
7: Temp. 4, Time 5& 240 & 250 & 260 & 270 & & 552 & 1008 & 2016 & 3528 & 5040 \\
8: Temp. 5, Time 5& 240 & 250 & 260 & 270 & 280 & 552 & 1008 & 2016 & 3528 & 5040 \\\hline\hline
\end{tabular}
\end{center}
\end{table}

\subsection{Results under the Correct Model}
\begin{figure}
\begin{center}
\includegraphics[width=.8\textwidth]{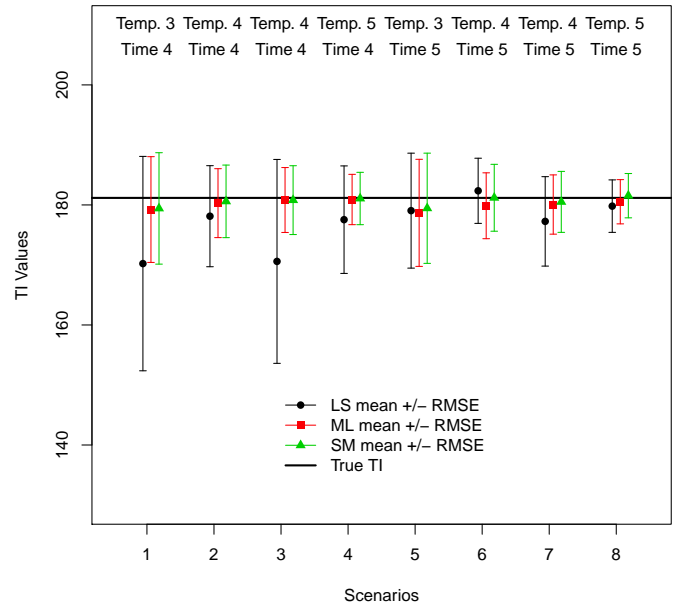}
\end{center}
\caption{Plot of the estimated mean, bias, SD, and RMSE of the TI estimators for the traditional method (TM), the parametric method (PM), and the semiparametric method (SPM) for Setting I: the parametric model is correctly specified.}\label{fig:sim.result.setting.I}
\end{figure}

Table~\ref{tab:sim.result.setting.I} shows the estimated mean, bias, standard deviation (SD), and root of mean squared error (RMSE) of the TI estimators for the traditional method (TM), the parametric method (PM), and the semiparametric method (SPM) for Setting I: the parametric model is correctly specified. Figure~\ref{fig:sim.result.setting.I} visualize the results in Table~\ref{tab:sim.result.setting.I}. We observe that those scenarios with more time points and temperature levels tend to have better precision in estimating TI for all methods. Testing at higher temperature levels tends to provide better precision for all the methods. Among the three methods, the traditional method tends to perform worse than the other two methods. This observation for the traditional method is consistent with the findings in King et al.~\cite{Kingetal2016}. The performance of the newly added semiparametric is comparable to the parametric method.

\begin{table}
\caption{Estimated mean, bias, SD, and RMSE of the TI estimators for the traditional method (TM), the parametric method (PM), and the semiparametric method (SPM) for Setting I: the parametric model is correctly specified.}\label{tab:sim.result.setting.I}
\begin{center}
\begin{tabular}{c|c|ccc|ccc|ccc|ccc}\hline\hline
\multirow{2}{*}{Scenarios} & \multirow{2}{*}{True TI} & \multicolumn{3}{c}{Mean} & \multicolumn{3}{|c|}{Bias} & \multicolumn{3}{|c|}{SD} & \multicolumn{3}{|c}{RMSE}\\\cline{3-14}
&  & TM & PM &SPM & TM & PM &SPM & TM & PM &SPM & TM & PM &SPM\\\hline
1: Temp. 3, Time 4& 181 & 170 & 179 & 179 & 11 & 2 & 2 & 14 & 9 & 9 & 18 & 9 & 9 \\
2: Temp. 4, Time 4& 181 & 178 & 180 & 181 &  3 & 1 & 1 &  8 & 6 & 6 &  8 & 6 & 6 \\
3: Temp. 4, Time 4& 181 & 171 & 181 & 181 & 11 & 0 & 0 & 13 & 5 & 6 & 17 & 5 & 6 \\
4: Temp. 5, Time 4& 181 & 178 & 181 & 181 &  4 & 0 & 0 &  8 & 4 & 4 &  9 & 4 & 4 \\\hline
5: Temp. 3, Time 5& 181 & 179 & 179 & 179 &  2 & 2 & 2 &  9 & 9 & 9 & 10 & 9 & 9 \\
6: Temp. 4, Time 5& 181 & 182 & 180 & 181 &  1 & 1 & 0 &  5 & 5 & 6 &  5 & 5 & 6 \\
7: Temp. 4, Time 5& 181 & 177 & 180 & 181 &  4 & 1 & 1 &  6 & 5 & 5 &  7 & 5 & 5 \\
8: Temp. 5, Time 5& 181 & 180 & 181 & 182 &  1 & 1 & 0 &  4 & 4 & 4 &  4 & 4 & 4 \\\hline\hline
\end{tabular}
\end{center}
\end{table}

\subsection{Results under a Misspecified Model}
\begin{figure}
\begin{center}
\includegraphics[width=.8\textwidth]{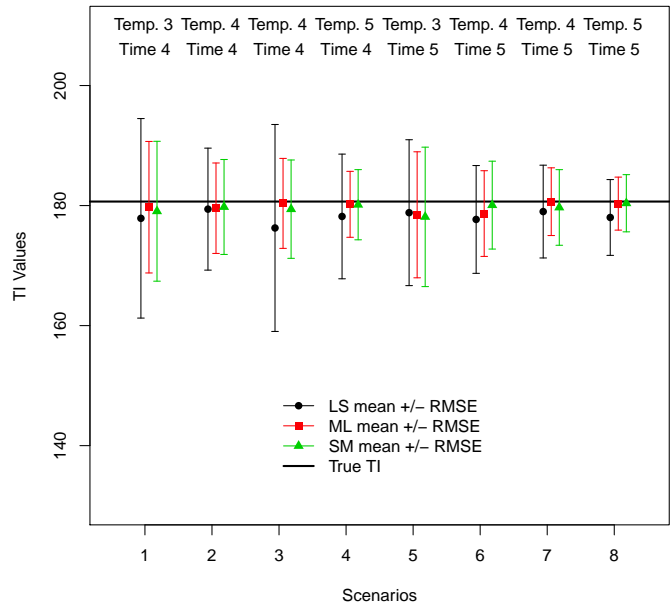}
\end{center}
\caption{Plot of the estimated mean, bias, SD, and RMSE of the TI estimators for the traditional method (TM), the parametric method (PM), and the semiparametric method (SPM) for Setting II: the parametric model is incorrectly specified.}\label{fig:sim.result.setting.II}
\end{figure}

Table~\ref{tab:sim.result.setting.II} shows the estimated mean, bias, SD, and RMSE of the TI estimators for the traditional method, the parametric method, and the semiparametric method for Setting II: the parametric model is incorrectly specified. Figure~\ref{fig:sim.result.setting.II} visualizes the results in Table~\ref{tab:sim.result.setting.II}. We observe similar patterns to Setting I. That is, those scenarios with more time points and temperature levels tend to have better precision in estimating TI for all methods, and the traditional method tends to perform worse than the other two methods, which is also consistent with the findings in King et al.~\cite{Kingetal2016}. Surprisingly the parametric method performs well even under model misspecification. Similarly, the performance of the newly added semiparametric is comparable to the parametric method.

\begin{table}
\caption{Estimated mean, bias, SD, and RMSE of the TI estimators for the traditional method (TM), the parametric method (PM), and the semiparametric method (SPM) for Setting II: the parametric model is incorrectly specified.}\label{tab:sim.result.setting.II}
\begin{center}
\begin{tabular}{c|c|ccc|ccc|ccc|ccc}\hline\hline
\multirow{2}{*}{Scenarios} & \multirow{2}{*}{True TI} & \multicolumn{3}{c}{Mean} & \multicolumn{3}{|c|}{Bias} & \multicolumn{3}{|c|}{SD} & \multicolumn{3}{|c}{RMSE}\\\cline{3-14}
&  & TM & PM &SPM & TM & PM &SPM & TM & PM &SPM & TM & PM &SPM\\\hline
1: Temp. 3, Time 4& 181 & 178 & 180 & 179 & 3 & 1 & 2 & 16 & 11 & 12 & 17 & 11 & 12 \\
2: Temp. 4, Time 4& 181 & 179 & 180 & 180 & 1 & 1 & 1 & 10 & 7 & 8 & 10 & 8 & 8 \\
3: Temp. 4, Time 4& 181 & 176 & 180 & 179 & 4 & 0 & 1 & 17 & 7 & 8 & 17 & 7 & 8 \\
4: Temp. 5, Time 4& 181 & 178 & 180 & 180 & 2 & 0 & 1 & 10 & 5 & 6 & 10 & 5 & 6 \\\hline
5: Temp. 3, Time 5& 181 & 179 & 178 & 178 & 2 & 2 & 3 & 12 & 10 & 11 & 12 & 10 & 12 \\
6: Temp. 4, Time 5& 181 & 178 & 179 & 180 & 3 & 2 & 1 & 8 & 7 & 7 & 9 & 7 & 7 \\
7: Temp. 4, Time 5& 181 & 179 & 181 & 180 & 2 & 0 & 1 & 8 & 6 & 6 & 8 & 6 & 6 \\
8: Temp. 5, Time 5& 181 & 178 & 180 & 180 & 3 & 0 & 0 & 6 & 4 & 5 & 6 & 4 & 5 \\  \hline\hline
\end{tabular}
\end{center}
\end{table}

\section{Discussions}\label{sec:discussions}
In literature, there are three methods available to estimate the TI based on ADDT data, which are the traditional method, the parametric method, and the semiparametric method.  In this chapter, we provide a comprehensive review of the three methods and illustrate how the TI can be estimated based on different models. We also conduct a simulation study to show the properties of different methods. The comparisons and discussion in this chapter can be useful for practitioners and future industrial standards.

Here, we provide a summary on the pros and cons of each method.

\begin{inparaitem}
\item Regarding estimation performance, if there are fewer number of temperature levels/number of time points, the traditional method tends to not performance well. When there are five temperature levels and five time points, the traditional method works well. Both the parametric and semiparametric methods perform better than the traditional methods and their performance are comparable to each other.

\item Regarding model assumption, the traditional method does not require specific forms for the underlying degradation path because it uses polynomial interpolation. The semiparametric method does not require a specific form but assume the underlying path is monotone and smooth. The parametric method assumes a specific form, which requires the strongest assumption. However, the simulation study show that the parametric model used here is flexible to some extent under model misspecification.

\item Regarding data use, both the parametric and semiparametric methods use all data for analyses, including those have not yet reached the failure threshold. The traditional method will discard the data from the temperature which has not reached the failure threshold yet.

\item Both the parametric and semiparametric methods can quantify the uncertainties in the estimation (see King et al.~\cite{Kingetal2016}, and Xie et al.~\cite{XieKingHongYang2015} for details). Because the traditional method requires two steps to estimate the TI, it is challenging to quantify the statistical uncertainties.

\item The semiparametric method is the most computationally intensive one, and the parametric method is in the middle in term of computational time. All the three methods is implemented in the R package ADDT. The chapter in \cite{JinXieHongVanMullekom2017} gives a detailed illustration for the use of the package.

\end{inparaitem}

In summary, it is of advantages to use the parametric and semiparametric methods in the ADDT analysis and TI estimation. In practice, one can compare the model fitting of both the parametric and semiparametric methods (e.g., AIC values) to determine which models can provide a better description to the ADDT data. The practitioner should also weigh the pros and cons discussed in this section in conjunction with the minimum AIC model for final model selection. Details of model comparisons can be found in Xie et al.~\cite{XieKingHongYang2015}.

\section*{Acknowledgments}\label{Acknowledgement}
The authors acknowledge Advanced Research Computing at Virginia Tech for providing computational resources. The work by Hong was partially supported by the National Science Foundation under Grant CMMI-1634867 to Virginia Tech.


\end{document}